\documentclass[aps,prb,twocolumn,superscriptaddress,
preprintnumbers,amsmath,amssymb
,floatfix
]{revtex4}

\usepackage{graphicx}        
\usepackage{mathrsfs} 

\usepackage{graphicx}
\usepackage{dcolumn}
\usepackage{bm}
\usepackage{amsmath}
\usepackage{amssymb}
\usepackage{revsymb}

\begin{document}

\affiliation{
Department of Physics and Astronomy, Georgia State
University, Atlanta, Georgia 30303, USA}

\title{Giant Surface Plasmon Induced Drag Effect (SPIDEr) in Metal Nanowires}


\author{Maxim Durach}
\affiliation{
Department of Physics and Astronomy, Georgia State
University, Atlanta, Georgia 30303, USA}
\author{Anastasia Rusina}
\affiliation{
Department of Physics and Astronomy, Georgia State
University, Atlanta, Georgia 30303, USA}
\author{Mark I. Stockman}
\affiliation{
Department of Physics and Astronomy, Georgia State
University, Atlanta, Georgia 30303, USA}
\affiliation{
Max Planck Institute for Quantum Optics,
Hans-Kopfermann-Strasse 1, 85748 Garching, Germany}
\affiliation{
Ludwig Maximilian University Munich, Am
Coulombwall 1, 85748 Garching, Germany}
\email{mstockman@gsu.edu}
\homepage{http://www.phy-astr.gsu.edu/stockman}

 

\date{\today}

\begin{abstract}
Here, for the first time we predict a giant surface plasmon-induced 
drag effect (SPIDEr), which exists under
conditions of the extreme nanoplasmonic confinement. 
Under realistic conditions, in nanowires, this giant SPIDEr generates
rectified THz potential differences up to 10 V and extremely strong
electric fields up to $\sim 10^5-10^6$ V/cm.
The SPIDEr is an ultrafast effect whose bandwidth
for nanometric wires is $\sim 20$ THz. The giant SPIDEr opens up a new
field of ultraintense THz nanooptics with wide potential applications in
nanotechnology and nanoscience, including microelectronics, nanoplasmonics, and
biomedicine.
\end{abstract}

\maketitle

The fact that electromagnetic fields exert mechanical forces on 
matter is well known
and has found applications in atomic physics%
\cite{Hansch_Schawlow_Opt_Comm_1975_laser_cooling,%
Letokhov_JETP_atom_cooling} and bio- and nano\-technology,%
\cite{
Ashkin_Science_1987_Traping_Viruses,Novotny_PRL_1997_Nanotweezers,%
Volpe_et_al_PRL_2006_SP_Forces} 
and picosecond photodetectors based on the photon drag effect (PDE).%
\cite{Ganichev_Intense_Terahertz_Excitation_of_Semiconductors_2006}
The semiconductor PDE detectors have proved to be very
practical for relatively fast detection of picosecond pulses 
in a wide frequency range
spanning from THz to infrared. 
It has beeen proposed to use enhanced fields in phonon-polariton
silicon carbide structures for laser particle accelerators.%
\cite{Shvets_et_al_Proc_Royal_Soc_2006_Laser_Acceleration}
There have been experimental investigations of the PDE 
in metal nanofilms that are thicker than the skin depth%
\cite{Ishihara_APL_2005_SP_enhanced_photon_drag,%
Ishihara_Opt_Expr_rectification_1d_photonic_crystal}
Metals support 
surface plasmon polaritons (SPPs)%
\cite{Bozhevlolny_Plasmonic_Nanoguides_Book}
that exert forces on electrons
causing an SPP-enhanced PDE.
\cite{Ishihara_APL_2005_SP_enhanced_photon_drag}
However, it is typically rather small, with the induced potential differences in
the mV range. This modest enhancement of the PDE in these experiments
was due to relatively slow variation of SPP fields in space, 
and the momentum transferred to the electrons was correspondingly small.
Because of the long relaxation times of the SPPs at plane
metal surfaces, this PDE is also expected to be relatively slow. 
A drag effect under the conditions of strong nanoplasmonic
confinement, when the SPP localization radius is less than the skin
depth ($\sim 25$ nm), has not been 
studied or exploited 
theoretically or experimentally.

In this Letter, we predict a giant 
surface plasmon-induced drag effect (SPIDEr) in metal nanowires,
which is fast, with response on the femtosecond time scale.
We show that the ultrashort, nanolocalized SPP pulses 
exert forces on electrons in the nanowires,
inducing giant THz electromotive force (emf) along the SPP propagation
direction.
We have found that in thin ($\sim 5$ nm radius) wires 
this emf can reach 
$\sim 10$ V, with nanolocalized THz fields as high as $\sim 1$ MV/cm.
Such THz field have previouisly been generated in the far zone%
\cite{Sell_Leitenstorfer_Huber_OL_2008_Ultraintense_THz}, where they
produce non-perturbative effects,%
\cite{Huber_et_al_PRL_2008_THz_Strong_Field_Excitation_in_Cuprates}
but not on the nanoscale.
In contrast, the plasmonic metal nanowires 
can serve as nanolocalized sources of high THz fields.
We also study the dynamics of the SPIDEr for short SPP pulses
and suggest that adiabatically tapered nanowires%
\cite{Phys_Rev_Lett_93_2004_Tapered_Plasmonic_Waveguides}
can be used as broadband nanoscopic photodetectors
with extremely fast response due to the femtosecond momentum relaxation
times in metals.%
\cite{Kruglyak_Momentum_Relaxation_Times_Metals_PRB_2005} The nature of
the giant enhancement of the SPIDEr is novel in nanoplasmonics: it is
not the enhancement of the optical fields {\it per se} (the maximum
magnitude of the local fields is limited by the breakdown at the metal
surface that occurs for fields $\sim 1$ V/\AA) 
but the very high {\it gradients} of these fields. 
The SPIDEr is ultrafast because
it is a non-resonant effect whose bandwidth is
comparable to that of the entire optical spectrum.

\begin{figure}
\centering
\includegraphics[width=.40\textwidth]
{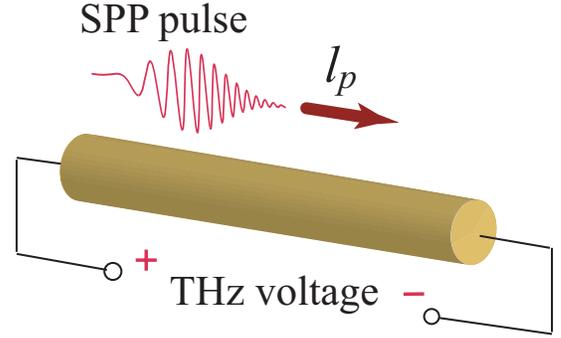}
\caption{\label{schematic_2.eps}
Schematic of SPIDEr in metal nanowire. Propagating SPPs
create forces acting on carriers in the nanowire, which
leads to THz-band voltage (emf) between the ends of the wire. Picosecond or
femtosecond pulses can be used to manipulate
the time dependence of the created emf.
}
\end{figure}

Consider a metal nanostructure propagating an SPP pulse.
The SPP field produces polarization $\mathbf P$, 
charges with macroscopic density 
$\rho=-\mathrm{div} \mathbf{P}$, current density 
$\mathbf{j}=\partial {\mathbf{P}}/\partial t$, and the 
surface charge density $\sigma=(\mathbf{P}\cdot\mathbf{n})$ 
at the surface of the metal,
where $\mathbf{n}$ is the normal to the surface pointing outward.
We do not consider systems with optical magnetism, whose introduction at
optical frequency is problematic%
\cite{Landau_Lifshitz_Electrodynamics_Continuous:1984,%
Merlin_PNAS_2009_Optical_Magnetism}
Therefore we set $\mathbf{B}=\mathbf{H}$, which precludes the existence
of surface currents.
Using Eqs. (\ref{Lforce_vol2})-(\ref{volume_int})  
of the Methods Section, 
we obtain the following  
general expression for the total force
\begin{equation}
\mathbf{F}=\int_V \left[\mathrm{grad}(\mathbf{P}^c\cdot \mathbf{E})%
+\frac{1}{c}\frac{\partial(\mathbf{P}\times\mathbf{B})}{\partial t}%
\right]dV~,~~ 
\label{total_force_res}
\end{equation}
where superscript ``$c$''
implies that the differentiation does not apply to the labeled vector.
This result is of fundamental importance for processes 
involving interaction of nanoplasmonic fields with metal electrons.
\footnote{We calculate {\it macroscopic} forces 
assuming that the nanostructure is homogeneous enough
and the frequencies under consideration are far enough from 
the surface plasmon (SP) resonances. 
Then the field fluctuations, 
characteristic of the disordered systems,%
\cite{Stockman:1994_Giant_Fluctuations}
could be disregarded. 
}
Equation (\ref{total_force_res}) is valid for a wide range of problems
with a general material equation $\mathbf{P}=\hat{\chi}\mathbf{E}$, 
including those where operator $\hat{\chi}$ describes anisotropic
or non-local media.
The first term in Eq.\ (\ref{total_force_res}) 
is similar to the force acting on a point dipole moment.
\cite{Shimizu_Sasada_AJP_1998_dipole_forces}
The second term in Eq.\ (\ref{total_force_res}) is the Abraham force.
In a monochromatic field, this force averaged over the period 
of oscillations is zero,
but in the field of a pulsed excitation it has a finite magnitude. 

We apply this fundamental Eq.\ (\ref{total_force_res})
to describe the SPIDEr. For certainty, 
consider a metal nanowire with radius $R$ and dielectric susceptibility 
$\chi(\omega)$, which is 
oriented along the $z$-axis and embedded into 
a dielectric with a dielectric permittivity of $\varepsilon_d$. 
This wire propagates an SPP pulse, which can be excited  
by external sources using, e.g., the effect of adiabatic compression.%
\cite{Phys_Rev_Lett_93_2004_Tapered_Plasmonic_Waveguides,%
Verhagen_Polman_Kuipers_Opt_Express_2008_Adiabatic_Nanofocusing}

In the case of extreme nanoplasmonic confinement ($R\ll l_s$,
where $l_s$ is the skin depth), $R$
becomes the only relevant quantity of the dimensionality of length.
\cite{Phys_Rev_Lett_93_2004_Tapered_Plasmonic_Waveguides} Therefore
there is scaling of all magnitudes in $R$. 
The SPP wave power $\mathscr P$ scales as $\mathscr P\sim E^2 R^2 v_g$,
where $v_g\sim \omega R$ is the SPP group velocity. The SPIDEr-induced
potential difference [electromotive force (emf)] $\mathscr E$ is
proportional to the pressure produced by force (\ref{total_force_res}),
$\mathscr E\sim F/R^2$. The propagation length of the SPP $l_p\sim R
Q$, where $Q$ is the SPP figure of merit, 
independent of $R$. 
These arguments allow us to predict scaling of the
SPIDEr force $F$, emf $\mathscr E$, and the electric field due to SPIDEr
rectification $E_R$ (which for femtosecond SPP pulses possesses THz
frequencies):
$
F\propto \mathscr P R^{-1}~,~~~ \mathscr E\propto \mathscr P R^{-3}~,~~~
E_R\propto \mathscr P R^{-4}~,~~~ E_{mR}\propto R^{-1}~.
$
We have also indicated the scaling of the maximum rectified field $E_{mR}$ 
(at the maximum tolerable power $\mathscr P_m$). 
The scaling implies that all the effects caused by the SPIDEr increase
with decreasing the wire radius as its powers. This enhancement is not
resonant and therefore has 
bandwidth comparable to that of the entire optical spectrum. 
The scaling 
describes only the dependence on $R$.
There are also prefactors describing the dependence on dielectric
permittivities, frequency, etc. They take into account 
an additional enhancement close to the SP resonant frequency,
which is multiplicative in magnitude. Below in this paper we show that 
this scaling 
is reproduced 
by the theory results. 

The SPPs are transverse magnetic (TM) modes, and their
complex fields have the form
\begin{equation}
\mathbf{E}=A(t^\prime)(\tilde E_{z}\hat{\mathbf{z}}+\tilde E_{\rho}\hat{\bm{\rho}})
e^{i(k z-\omega t)},~
\mathbf{H}=A(t^\prime)\tilde H_{\varphi}\hat{\bm{\varphi}}e^{i(k z-\omega t)},~
\label{SPP_fields} 
\end{equation}
where $t^\prime=t-z/v_g$, and $v_g$ is the SPP group velocity at
the pulse carrier frequency $\omega$, $k$ is the SPP wave number.%
\cite{Phys_Rev_Lett_93_2004_Tapered_Plasmonic_Waveguides}
The total power flowing through the plane $z=0$ at the moment $t$ 
is $\mathscr{P}(t)=2\pi\int_0^\infty %
\rho d\rho \bar{S}_z(\mathbf{r},t)|_{z=0}$,
where $\bar{\mathbf{S}}(\mathbf{r},t)=%
(c/8\pi)\mathrm{Re}[\mathbf{E}\times\mathbf{H}^\ast]$
is the Pointing vector averaged over SPP period. 
Considering the azimuthally-symmetric (TM$_0$) modes, 
functions $\tilde E_z$, $\tilde E_\rho$ and $\tilde H_\varphi$ 
depend only on radius $\rho$. 
We normalize them for real amplitude $A(t)$ to
satisfy a relation $A^2(t)=\mathscr{P}(t)$.
Disregarding the group velocity dispersion in Eq.\
(\ref{SPP_fields}) 
is valid for pulses with duration of tens of femtoseconds and
greater, and frequencies not too close to the SP resonance.

The momentum transferred from the electromagnetic field to 
the electronic system implies forces exerted on the electrons. The
density of these forces are given by Eqs.\ (\ref{Lforce_vol}) and 
(\ref{Lforce_surf}) of the Methods Section. 
These forces lead to an emf, which corresponds to 
optical rectification in the system, which for femtosecond SPP pulses 
results in the emf in the THz frequency range. 
Since the electron momentum-relaxation 
time is on the scale of femtoseconds, electrons come to a
local equilibrium in the process of this rectification. Therefore,
we will describe it in the hydrodynamic approximation, for which the pressure
$p$ and electrostatic potential $\phi$ satisfy an equation
$p+ne\phi=const$, where $e$ is electron charge, and $n$ is electron
density. From this equation, we can find the emf $\mathscr E=\Delta \phi$, 
which is the total change of potential 
in the direction of SPP propagation (the $z$
direction), $\mathscr E=-\Delta p/(ne)$, where 
$\Delta p=\bar{F}_z/(\pi R^2)$ is
the full change of the pressure. 
Here, $\bar{F}_z$ is $z$-component 
of the force\ (\ref{total_force_res}) averaged over the period of SPP
oscillations. 

The total force $\bar{F}_z$ 
is composed of three forces 
(see Eq. (\ref{Total_Force}) in Methods Section): 
the SPP pressure, striction, and Abraham force 
$\bar{F}_z=f_z^{pr}+f_z^{st}+f_z^A$.
These three forces result in 
three terms of the SPIDEr emf
\begin{equation}
\mathscr{E}=R_H
\left(\frac{\mathscr{F}(t)}{A_{pr}}+
\frac{ \mathscr{P}(t)}{A_{st}}+
\frac{\mathscr{F}^\prime(t)}{c L_{A}}\right)~,~~
\label{potential}
\end{equation}
where $R_H=-1/(ecn)$ is the Hall constant, and 
coefficients $A_{pr}$, $A_{st}$, and $L_A$ and $\mathscr F(t)$ are 
defined by Eqs.\ (\ref{func})-(\ref{A_Abraham_force}) of 
the Methods Section.
The Abraham force contribution [the third term in Eq.\
(\ref{potential})] is small under a condition $k_0 l_p / (k c
\tau)\ll 1$, where $k_0=\omega/c$, and $\tau$ is the pulse duration. This
condition is well satisfied for the parameters used in this paper.
We will consider two limiting cases pertaining to Eq.\
(\ref{potential}): 
a quasi-monochromatic regime of long pulses ($\tau\gg t_p$) and 
a regime of short pulses ($\tau\ll t_p$), where $t_p$
is the SPP dissipation time.
Consider first the quasi-monochromatic regime where 
$\mathscr{F}(t)\approx \mathscr{P}(t)$ 
(see Eq.\ (\ref{func}) of the Methods Section). 
In such a case, the total emf $\mathscr{E}(t)$ 
follows the pulse-envelope time dependence
$\mathscr P(t)$. 
The computations will be made for a silver%
\cite{Johnson:1972_Silver} nanowire embedded in vacuum ($\varepsilon_d=1$).

\begin{figure}
\centering
\includegraphics[width=.48\textwidth]
{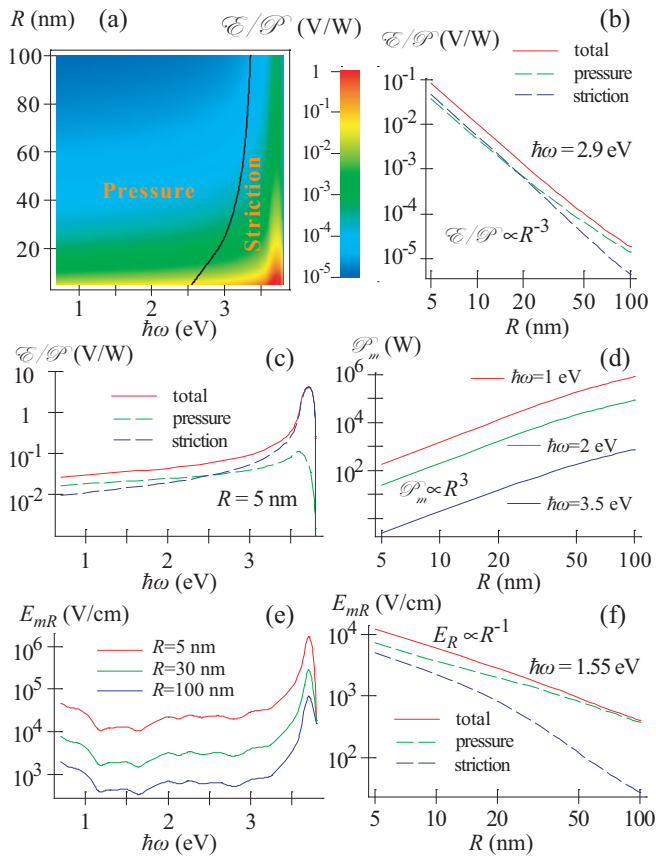}
\caption{\label{dc.eps}
SPIDEr for quasi-monochromatic SPP pulses: emf and
rectified field dependence on the
frequency $\hbar\omega$ and wire radius $R$. Note the logarithmic
scale for the magtnitude of the effect.
(a) Dependence of the SPIDEr emf per unit SPP power $\mathscr{E/P}$ on 
wire radius and frequency. The black broken curve indicates the
parameters at which SPP pressure is equal to striction. 
The magnitude of the effect is denoted by the color-coding bar. 
(b) Dependence of SPIDEr magnitude $\mathscr{E}$ on wire radius $R$
per unit power of the SPP wave 
(solid red curve). 
The contributions of the pressure and striction to the total 
magnitude of SPIDEr are shown by the dashed curves. 
(c) Dependence of SPIDEr magnitude per unit power $\mathscr{E/P}$ 
on frequency $\omega$ for $R=5$ nm.
(d) Maximum power that a wire can tolerate $\mathscr P_m$ as a function
of wire radius $R$ for different frequencies $\omega$.
(e) The maximum SPIDEr rectified field $E_{mR}$ (for the maximum tolerable
power $\mathscr{P}_m$) as a function of frequency for three wire radii
$R=5$, 30, and 100 nm. 
(f) The maximum SPIDEr rectified field $E_{mR}$ (for the maximum tolerable
power $\mathscr{P}_m$) as a function of the wire radius $R$ for
frequency $\hbar\omega=1.55$ eV.
}
\end{figure}

For the quasi-monochromatic case, the total emf $\mathscr{E}$ as a
function of frequency $\omega$ and wire radius $R$ is displayed in 
Fig.\ \ref{dc.eps} (a). In contrast to the case of dielectric media, 
in the present plasmonic case the 
SPP pressure and striction contributions to the 
emf have the same sign since $\chi^\prime<0$.  These two
contributions are equal if a condition
$\chi^{\prime\prime} Q =-\chi^\prime$ is satisfied. 
The black solid line represents this condition; to the left of this line
the pressure dominates, and to the right the striction force gives the
major contribution to the SPIDEr. This is understandable because close
to the SP resonance of the wire (at $\approx 3.7$ eV), the gradient of
the SPP intensity increases due to the high loss: the striction force is
of a gradient nature, dominating therefore. Similarly, with the decrease
of $R$, the intensity gradient increases due to the increased
confinement, which also leads to the relative increase of the striction
with respect to the pressure force, as we clearly can see from this and
other panels of Fig.\ \ref{dc.eps}. General increase of the SPIDEr 
at the SP resonant frequency can be seen as a broad red peak.

Magnitude of the SPIDEr emf relative to the SPP wave power,
$\mathscr{E/P}$, is illustrated in Fig.\ \ref{dc.eps} (b) as a function of
the wire radius $R$ for a frequency of $\hbar\omega=2.9$ eV. 
The SPIDER effect is gigantically
enhanced for strong nanoplasmonic confinement: by four orders of
magnitude when $R$ decreases from 100 to 5 nm at the same SPP power.
There is a pronounced scaling $\mathscr{E/P}\propto R^{-3}$ at
$R\lesssim l_s$, in accord with the discussion following
Eq. (\ref{total_force_res}).

The spectral dependence
of the relative SPIDER emf, $\mathscr{E/P}$, for a wire of the smallest radius 
considered, $R= 5$ nm, is depicted in Fig.\ \ref{dc.eps}(c).%
\footnote{For significantly thinner wires, nonlocal-response effects may
become significant, cf. Refs.\ 
\onlinecite{Larkin_Stockman_Nano_Letters_5_339_2005_Imperfect_Perfect_Lens,%
Aizpurua_Rivacoba_PRB_2008_Nonlocal_Plasmons_in_Nanowires}
}
Importantly, the magnitude of the emf in this case is very large, 
from 0.01 to 10 V per 1 W of the SPP pulse power, in
the entire optical range, with a pronounced resonance at the SP
frequency.
This large magnitude shows that the SPIDEr effect
can be used for the photodetection on the nanoscale, i.e., in the role
that previously was deemed only available for semiconductors. In this
sense, it belongs to the area of the active
nanoplasmonics.%
\cite{MacDonald2009_et_al_Nature_Photonics_3_55_2009_Ultrafast_Active_Plasmonics}

By classification of the nonlinear optics, the SPIDEr is an optical
second-order nonlinear effect: the magnitude of the SPIDEr emf 
$\mathscr E$ is proportional to the power $\mathscr P$ of
the SPP field. Therefore, the maximum achievable magnitude of the emf is
determined by the maximum $\mathscr P$ that the wire can tolerate. This
we estimate setting the optical field $E$ at the surface of the wire
equal to 1 V/\AA.%
\cite{Gault_APL_2005_strong_field_pulses_evaporation,%
Sha_APL_2008_strong_field_pulses_evaporation}
For fields significantly higher than this, there will
be massive ionization and damage of the metal surface. We plot in Fig.\
\ref{dc.eps} (d) this maximum intensity as a function of the wire radius
for three SPP frequencies. 
Note a very good scaling $\mathscr P_m\propto R^3$ 
in the region $R\lesssim l_s$ of the strong
nanoplasmonic confinement. These values of the $\mathscr P_m$ in
comparison to the data of Fig.\ \ref{dc.eps} (a)-(c) show that the gigantic 
values of of the SPIDEr emf $\mathscr E\sim 10$ V are realistically
achievable, which are many orders of magnitude greater than observed
previously in the metal films.
\cite{Ishihara_APL_2005_SP_enhanced_photon_drag,%
Ishihara_Opt_Expr_rectification_1d_photonic_crystal}

One of the most important for applications properties of the giant
SPIDEr is high local electric field $E_R$ generated due to the
SPIDEr optical rectification in the vicinity of the nanowire. Such a
field (averaged over the SPP decay length $l_p$)
can be found as $E_{R}=\mathscr E/l_p$. 
We display the maximum
achievable rectified field $E_{mR}$ (at the propagating SPP power of
$\mathscr P_m$) in Figs.\ \ref{dc.eps} (e), (f). As we see from panel (e),
the spectral dependence of the SPIDEr rectified THz field is very
similar for all wire sizes, but the magnitude of this field is much
greater for the 5 nm wire:
$E_{mR}\sim 10^5-10^6$ V/cm. The
nanolocalized THz fields of such a magnitude will excite a wealth of
nonlinear THz responses at the nanoscale.

\begin{figure}
\centering
\includegraphics[width=.48\textwidth]
{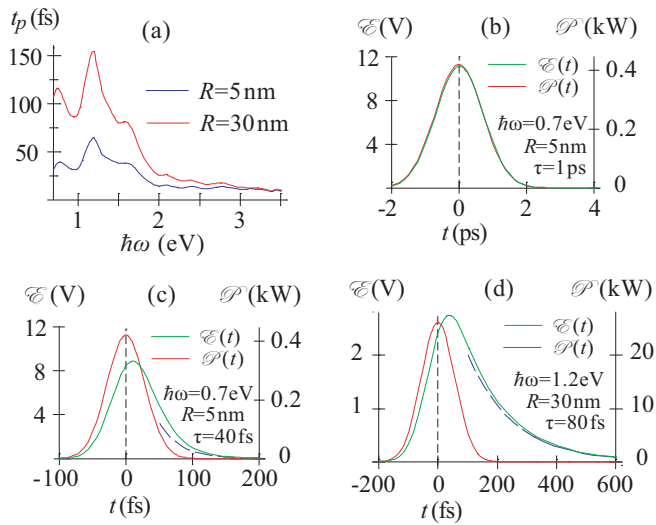}
\caption{\label{pulses.eps}
SPIDEr created by ultrashort SPP pulses:
fast femtosecond emf response.
(a) The dependence of the SPP lifetime $t_p=l_p/v_g$
on the frequency $\hbar \omega$ for $R=5~\mathrm{nm}$ and $R=30~\mathrm{nm}$.
(b) The time dependence of the emf $\mathscr E(t)$
(green line, left scale) and input power $\mathscr P(t)$ 
(red line, right scale).
The pulse duration is $\tau=1~\mathrm{ps} \gg t_p\approx 30~\mathrm{fs}$ 
and the emf closely follows the SPP pulse dynamics.
(c) The same for much shorter pulse with $\tau=40~\mathrm{fs}$.
The pressure-induced emf leads to small broadening in emf dynamics
(green line). The limiting exponential decay is outlined by the
broken blue line.
(d) Emf induced by the short pulse in nanowire with $R=30~\mathrm{nm}$
with frequency $\hbar\omega=1.2~\mathrm{eV}$. 
The emf response is broadened, since $\tau=80~\mathrm{fs}$, 
while $t_p\approx 150~\mathrm{fs}$.
}
\end{figure}

Now let us consider dynamics of the emf response to SPP pulses that
differ in duration $\tau$ with respect to the 
the SPP pulse dissipation time $t_p$. The latter is displayed in Fig.\
\ref{pulses.eps} (a) as a function of the frequency $\omega$.
As we can see 
time $t_p$ is in the range from 10 to 150 fs. The temporal
dependencies of the emf in comparison with the power $\mathscr P$
of the SPP pulses for various pulse durations is illustrated in
Figs.\ \ref{pulses.eps} (b)-(d). 
For a relatively long pulse ($\tau=1~\mathrm{ps}\gg t_p$)
shown in Fig.\ \ref{pulses.eps} (b),
the shape of the emf $\mathscr E(t)$ repeats that of the power 
$\mathscr P(t)$. This 
relatively long, picosecond response, nevertheless, corresponds to a 1
THz bandwidth for this nanowire used as a nanoscale photodetector. 
Note that the amplitude of the emf is very large, $\sim 10$ V.

For a much shorter, $\tau=40$ fs, SPP pulse and the same 5 nm wire, as
shown in Fig.\ \ref{pulses.eps} (c), there is a small broadening
and delay of the voltaic response (emf) $\mathscr E(t)$ with respect to
the excitation SPP pulse $\mathscr P(t)$. This broadening is due to the
pressure force that decays exponentially for long times, 
as Eq.\ (\ref{func}) of the Methods Section 
suggests, and the broken blue line in the figure indicates.
However, under the conditions considered, this delay and broadening are
not large. The frequency-response bandwidth of this wire
as an SPP photodetector on the nanoscale is very large, $\approx 20$
THz, which is characteristic of the extreme nanoplasmonic confinement.
The amplitude of the SPIDEr emf is also very large, 
$\mathscr E\sim 10$ V.

For a much thicker nanowire of $R=30$ nm (weak plasmonic confinement
case) and $\hbar\omega=1.2$ eV, illustrated in Fig.\ \ref{pulses.eps}
(d), the SPP decay time becomes much longer ($t_p=150$ fs). This leads
to a very significant delay and temporal broadening of the emf response
with a pronounced exponential part due to the pressure forces shown by
the broken blue line. These behavior is due to the much longer SPP
lifetimes for the weak confinement where a significant fraction of the
SPP energy propagates in the dielectric (vacuum). Nevertheless 
the emf response bandwidth is still very large,
on the order of 5 THz, and its amplitude is also very large, $\mathscr
E\gtrsim 1$ V.

In conclusion, the ultrafast giant SPIDEr in metal nanowires excited by
ultrashort SPP pulses is predicted in this Letter to generate a
gigantic emf up to $\sim 10$ V for the SPP waves of realistic 
and tolerable amplitudes. The SPIDEr
enhancement is mostly nonresonant, due to nanoplasmonic confinement of
the SPP fields, which leads to higher gradients of the
fields and their higher magnitudes. 
Because of its nonresonant nature,
the SPIDEr is an extremely fast effect:
frequency bandwidth of the generated THz fields is realistically 5-20
THz. Due to high longitudinal localization of the SPP waves in
the case of the strong nanoplasmonic confinement, the SPIDEr generates
very high local THz electric fields 
at the metal surface, $E_R\sim 10^5-10^6$ V/cm. Such fields are
capable of inducing strongly nonlinear responses, including dissociation
of molecules. Among possible applications of the
giant SPIDEr are rectification and detection of the nanoscale
femtosecond optical fields, coupling of the nanoplasmonic elements to
semiconductor devices, nonlinear THz spectroscopy on the nanoscale of
chemical and biological nanoobjects for biomedicine, etc.

This work was supported by grants from the Chemical Sciences,
Biosciences and Geosciences Division of the Office of Basic Energy
Sciences, Office of Science, U.S. Department of Energy, a grant
CHE-0507147 from NSF, and a grant from the US-Israel BSF. MIS work at
Garching was supported under contract from Ludwig Maximilian University of
Munich (Germany) in the framework of Munich Center of Advanced Photonics
(MAP). MIS gratefully acknowledges 
stimulating discussions with F. Capasso regarding photoinduced potentials 
in near-field optical microscope tips.

\section*{Methods}
Then  a macroscopic (averaged) Lorentz force density 
inside the material volume is given by
\begin{equation}
\mathbf{f}^v=-\mathbf{E}(\nabla\cdot\mathbf{P})+
\frac{1}{c}\frac{\partial \mathbf{P}}{\partial t}\times\mathbf{B}~.~~
\label{Lforce_vol}
\end{equation}
The surface polarization charges experience the Lorentz force with a
surface density
\begin{equation}
\mathbf{f}^s=\sigma \mathbf{E}~.~~
\label{Lforce_surf}
\end{equation}
The volume force density given by Eq.\ (\ref{Lforce_vol})
can be represented in the following way
\begin{equation}
\mathbf{f}^v=-\mathrm{div}\left(\mathbf{P}\otimes \mathbf{E}\right)
+\mathrm{grad}(\mathbf{P}^c\cdot \mathbf{E})
+\frac{1}{c}\frac{\partial(\mathbf{P}\times\mathbf{B})}{\partial t}~,~~
\label{Lforce_vol2}
\end{equation}
where $\otimes$ denotes the outer product of vectors.
Thus, the total force acting on the polarization charges in the SPP field is
\begin{equation}
\mathbf{F}=\int_V \mathbf{f}^v~dV+\oint_S \mathbf{f}^s~ds~,~~ 
\label{total_force}
\end{equation}
where the first integration runs over volume $V$ of the structure,
and the second is over its surface $S$.
Noticing that the volume integral of the 
first term in Eq.(\ref{Lforce_vol2}) is equal to
\begin{equation}
-\int_V \mathrm{div}\left(\mathbf{P}\otimes \mathbf{E}\right) dV=
-\oint_S (\mathbf{P}\cdot\mathbf{n})\mathbf{E}~ds=-\oint_S \mathbf{f}^s~ds~,~~
\label{volume_int}
\end{equation}
we cancel terms in Eq.\ (\ref{total_force}) to simplify it and obtain
the required Eq. (1) of the paper.

From Eq. (1) of the paper, 
the total force acting on electrons can
be expressed as a sum of the SPP pressure, striction, and Abraham force,
whose $z$-components are, correspondingly,
\begin{equation}
f_z^{pr}=\frac{\pi R^2 \mathscr{F}(t)}{c A_{pr}}~,~~
f_z^{st}=\frac{\pi R^2 \mathscr{P}(t)}{c A_{st}}~,~~
f_z^{A}=\frac{\pi R^2 \mathscr{F^\prime}(t)}{c^2 L_A}~,
\label{Total_Force}
\end{equation}
where
\begin{eqnarray}
&\displaystyle
\mathscr{F}(t)=\frac{1}{l_p}
\int_0^\infty \mathscr{P}(t^\prime)e^{-z/l_p} dz=~~~~~~~~~~~~~&\nonumber \\
&\displaystyle
=\frac{e^{-t/t_p}}{t_p}\int_{-\infty}^t 
\mathscr{P}(t^\prime)e^{t^\prime/t_p} dt^\prime~.~~
\label{func}
\end{eqnarray}
In Eq.\ (\ref{Total_Force}),
coefficients $A_{pr}$ and $A_{st}$ have dimensionality of area, and $L_A$
has dimensionality of length. They are defined as
\begin{eqnarray}
&\displaystyle
A_{pr}=\left(\frac{c\chi^{\prime\prime}Q}{2R^2} 
~\int_0^R |\mathbf{\tilde E}|^2 \rho d\rho\right)^{-1}~,~~
& \label{A_SPP_pressure} \\
&\displaystyle
A_{st}=- \left(\frac{c\chi^\prime}{2R^2}%
\int_0^R |\mathbf{\tilde E}|^2 \rho d\rho\right)^{-1}~,~~ 
&\label{A_SPP_striction} \\
&\displaystyle
L_A=\left(\frac{c l_p}{R^2}%
\int_0^R \mathrm{Re}\left(\chi \tilde E_\rho \tilde H_\varphi^\ast\right)
\rho d\rho\right)^{-1}~,~~
&\label{A_Abraham_force}
\end{eqnarray}
where $Q = \mathrm{Re}k/\mathrm{Im}k$ is the SPP figure of merit,
$|\mathbf{\tilde E}|^2=|\tilde E_\rho|^2+|\tilde E_z|^2$, 
$l_{p}=1/(2\mathrm{Im}\,k)$ is 
the SPP propagation length, and
$t_p=l_p/v_g$ is the SPP pulse lifetime.






\end{document}